\newcommand{\Red}[1]{\textcolor[rgb]{1.00,0.00,0.00}{#1}}
\newcommand{\Blue}[1]{\textcolor[rgb]{0.00,0.00,1.00}{#1}}

\documentclass[sigconf, screen]{acmart}
\acmConference[ICSE 2022]{The 44th International Conference on Software Engineering}{May 21–29, 2022}{Pittsburgh, PA, USA}
\usepackage{graphicx}
\usepackage{textcomp}
\usepackage{xcolor}
\usepackage{color}
\usepackage{natbib}
\setcitestyle{square, comma, numbers,sort&compress, super}
\usepackage{amsmath,amsfonts}
\usepackage{pifont}
\usepackage{multirow}
\usepackage{algorithm}  
\usepackage{algpseudocode}  
\usepackage{amsmath}  
\usepackage{balance}
\usepackage{subcaption, booktabs}

\newcommand{\cmark}{\ding{51}}%
\newcommand{\xmark}{\ding{55}}%

\copyrightyear{2022} 
\acmYear{2022} 
\setcopyright{acmcopyright}\acmConference[ICSE '22]{44th International Conference on Software Engineering}{May 21--29, 2022}{Pittsburgh, PA, USA}
\acmBooktitle{44th International Conference on Software Engineering (ICSE '22), May 21--29, 2022, Pittsburgh, PA, USA}
\acmPrice{15.00}
\acmDOI{10.1145/3510003.3510158}
\acmISBN{978-1-4503-9221-1/22/05}

\begin{document}


\title{ARCLIN: Automated API Mention Resolution for Unformatted Texts}

\author{Yintong Huo}
\affiliation{%
  \institution{The Chinese University of Hong Kong}
  \city{Hong Kong}
  \country{China}}
\email{ythuo@cse.cuhk.edu.hk}

\author{Yuxin Su}
\affiliation{%
  \institution{School of Software Engineering\\ Sun Yat-sen University}
  \city{Zhuhai}
  \country{China}}
\email{suyx35@mail.sysu.edu.cn}
\authornote{Corresponding author (suyx35@mail.sysu.edu.cn).}

\author{Hongming Zhang}
\affiliation{%
  \institution{The Hong Kong University of Science and Technology}
  \city{Hong Kong}
  \country{China}}
\email{hzhangal@cse.ust.hk}

\author{Michael R. Lyu}
\affiliation{%
  \institution{The Chinese University of Hong Kong}
  \city{Hong Kong}
  \country{China}}
\email{lyu@cse.cuhk.edu.hk}



\begin{abstract}

Online technical forums (e.g., StackOverflow) are popular platforms for developers to discuss technical problems such as how to use a specific Application Programming Interface (API), how to solve the programming tasks, or how to fix bugs in their code.
These discussions can often provide auxiliary knowledge of how to use the software that is not covered by the official documents.
The automatic extraction of such knowledge may support a set of downstream tasks like API searching or indexing.
However, unlike official documentation written by experts, discussions in open forums are made by regular developers who write in short and informal texts, including spelling errors or abbreviations.
There are three major challenges for the accurate APIs recognition and linking mentioned APIs from unstructured natural language documents to an entry in the API repository:
(1) distinguishing API mentions from common words; (2) identifying API mentions without a fully qualified name; and (3) disambiguating API mentions with similar method names but in a different library.
In this paper, to tackle these challenges, we propose an ARCLIN tool, which can effectively distinguish and link APIs without using human annotations.
Specifically, we first design an API recognizer to automatically extract API mentions from natural language sentences by a Conditional Random Field (CRF) on the top of a Bi-directional Long Short-Term Memory (Bi-LSTM) module, then we apply a context-aware scoring mechanism to compute the mention-entry similarity for each entry in an API repository.
Compared to previous approaches with heuristic rules,
our proposed tool without manual inspection outperforms by 8\% in a high-quality dataset Py-mention, which contains 558 mentions and 2,830 sentences from five popular Python libraries.
To our best knowledge, ARCLIN is the first approach to achieve full automation of API mention resolution from unformatted text without manually collected labels.
\end{abstract}





\maketitle

\section{Introduction}\label{section:introduction}
Application Programming Interface (API) is an essential component for programming. Developers use APIs to interact with a programming language or a software library. However, as a library contains thousands of APIs (e.g., PyTorch v1.8 has over 2,400 APIs) and there are hundreds of popular libraries in a language, it is impossible for developers to be familiar with all APIs. Therefore, developers are used to discussing programming-related questions in the online technical forum when they face troubles in programming tasks. One of the most popular forums, StackOverflow, contains over 20 million questions and 14 million users\footnote{The data dump is retrieved in September $1^{st}$, 2021.}.
It motivates researchers to explore how to identify knowledge in open forums to assist developers in many aspects, such as API recommendation~\cite{xie2020api}, API misuse detection~\cite{ren2020api, ren2020demystify}, and document augmentation~\cite{treude2016augmenting}.

The foundation of the above tasks is recognizing and identifying API mentions from an unstructured natural language. Conventionally, researchers tried to use rule-based methods to solve the task.
For example,~\citet{bacchelli2010linking,treude2016augmenting} identified API elements in texts by a set of regular expressions. ~\citet{huang2018api} chose a hyperlink in each StackOverflow post and used regular expressions to detect API entities. They also analyzed whether the text in HTML $<$code$>$ tag can match the API names in the API repositories. \citet{li2018api} detected APIs by checking whether the token of a sentence can match or partially match the name of an API by conducting minor modifications.~\citet{ren2020api} kept API mentions only in HTML $<$code$>$ elements.

However, these rule-based methods do not consider the short and informal nature of forum discussions, falling short in mining APIs in certain scenario. Typically, a forum may contain a large number of unprofessional developers with different technical backgrounds, who share the knowledge and information in their own writing styles. As a result, the API mentions could be in different formats. For example, previous study~\cite{tabassum-etal-2020-code} concluded from StackOverflow posts that 47\% of the API elements are not included with the HTML $<$code$>$ tag. Such inconsistency causes different kinds of ambiguity when we recognize and identify APIs. In this paper, we categorize these ambiguities into the following three types.


\begin{table*}[t]
    \small
    \centering
        \caption{Three main challenges for API mining in unformatted texts, \Blue{\textbf{Blue}} words refers to API mentions and \Red{\textbf{Red}} words refers to common words.}
    \begin{tabular}{l||l|l}
    \toprule
        Question ID & Sentence & API\\
    \midrule
       \#66952125  & So far I managed to use \Blue{\textbf{view}} once in my first very simple project... & torch.view()\\
       \#59905234 & You can use .numpy() to \Red{\textbf{view}} the internal data... & None\\
    \midrule
        \#42233297 & Simply reshaping the by \textbf{\Blue{np.reshape(data,(5000,3,32,32))}} would not work. &numpy.reshape()\\
        \#41518351 & The \textbf{\Blue{.reshape()}} method (of ndarray) returns the reshaped array. &numpy.reshape()\\
        \#47477945 & I have tried \textbf{\Blue{a.reshape(3,4)}} in for the numpy array but nothing is producing what I want.&numpy.reshape()\\
        \#65103822 & I would like to understand the difference between \Blue{\textbf{conv2d}} and \Blue{\textbf{conv3d}} in PyTorch. & torch.nn.Conv2d, torch.nn.Conv3d\\
    \midrule
        \#47532162 & I want to use the keras layer \Blue{\textbf{Flatten()}} or Reshape((-1,)) at the end of my model... & tensorflow.keras.layers.Flatten()\\
        \#60115633 & But in PyTorch, \Blue{\textbf{flatten()}} is an operation on the tensor. & torch.Tensor.flatten()\\
    \bottomrule
    \end{tabular}
    \label{tab:Challenge-example}
\end{table*}

The first one is \textit{common-word ambiguity}, referring to the ambiguity between common words and API mentions~\cite{ye2018apireal}. Traditionally, API name is composed of punctuations, brackets, and upper case letters; however, sometimes developers only write the API's method name in their answers, causing the difficulty of distinguishing it from common words. The first group in the Table~\ref{tab:Challenge-example} illustrates examples of this problem. Even if two sentences are all mentioning the word \textit{view}, the first sentence use \Blue{\textit{view}} to refer to the API \textit{torch.view()} whereas the second one use \Red{\textit{view}} as a common verb. Regular expressions fail in discriminating such API mentions with common words. Previous work~\cite{ye2018apireal} revealed that 35.1\% of the token $apply$ in StackOverflow posts tagged with Pandas actually referred to an API mention.

The second one is \textit{morphological ambiguity}, which is because developers rarely write down the full API name that can be perfectly matched with an API name in the library. Research on StackOverflow~\cite{chen2017unsupervised} concludes that morphological mentions, which include abbreviations, synonyms, and misspellings, are quite often in informal discussions. Four examples in the second group of Table~\ref{tab:Challenge-example} demonstrate the morphological variations. 
In the first three sentences, the API \textit{numpy.reshape()} was mentioned by replacing \textit{numpy} with its abbreviation \textit{np}, omitting library name, and using the customized variable name, respectively. 
The fourth sentence talks about the \textit{torch.nn.Conv2d} and \textit{torch.nn.Conv3d} APIs, but includes neither the library/module/class name nor the correct case (i.e., use \textit{conv2d} instead of \textit{Conv2d}). 


The third type is \textit{reference ambiguity}, which happens if the API lists contain various third-party libraries. The third group in Table~\ref{tab:Challenge-example} provides two instances of this problem. Even if both PyTorch library and Tensorflow library contain the API method \textit{flatten()}, we could characterize what the mentions refer to based on their sentence contexts (i.e., the first sentence mentions \textit{``keras''} module whereas the second one mentions \textit{``PyTorch''}). It is often the case that developers do not explicitly point out the specific library in their mentions, but such information can be derived from other words in the context.

Due to the above ambiguities, traditional information retrieval techniques cannot be effectively employed.
~\citet{dagenais2012recovering} applied a set of filtering heuristics to tackle the second challenge, but they failed in resolving common words ambiguity due to the shortcoming of regular expressions. 
The above challenges become more difficult if we apply the API mining task into unformatted sentences. Such free text does not contain any $<$code$>$ tags, so detecting API in this scenario is even harder. However, it is a non-negligible problem, since, in other scenarios (e.g., emails), we cannot use HTML tags. To make our research applicable for a broader application, we focus on mining APIs from free text. Although the most recent work \cite{ye2018apireal} claimed to distinguish API mentions from common words, they stored $<$code$>$ tags and code snippets in $<$pre$>$ $<$code$>$ tags from StackOverflow posts, instead of mining from free texts. Thus, their approach cannot be extended to general scenarios.

In this paper, to overcome the aforementioned ambiguity challenges, we propose a new API mining approach named ARCLIN (\textbf{A}PI \textbf{R}ecognition and \textbf{C}ontextual \textbf{LIN}king), which recognizes and identifies API mentions from natural language descriptions to a set of APIs without any human-annotated labels or handcrafted rules.
Our model is made up of an API recognizer that finds API mentions in free texts, and a contextual API linker that links API mentions to the correct API they refer to. 
Specifically, our API recognizer extensively deals with the first common word ambiguity by considering the context information in sentence-level around an API mention. 
For the words that are predicted to be an API mention, a library predictor inside the API linker predicts the related library to the sentence, restricting ARCLIN to link APIs in the predicted library, which resolves the reference ambiguity.
The similarity function in the API linker compares API mention with every entry in the API repository, considering both spelling similarity and lexical similarity, so minor morphological changes will not affect the linking result.
To the best of our knowledge, ARCLIN is the first approach that can automatically cope with these challenges above. 

Considering the numerous number of APIs in the real world, it is impractical to ask annotators to label such a large scale of data. To avoid this labor-intensive process, we design ARCLIN to be free from any human annotation in the training process by exploiting natural labels in the training set. 
Unlike human-labeled data, the automated labels may contain errors, but our API linker in the next step provides a strict selection to address this problem.
To evaluate the effectiveness of ARCLIN, we annotate a test set, which contains 2,948 sentences with 563 mentions under five popular third-party libraries.
On average, ARCLIN achieves 78.26\%, 73.53\% and 75.82\% in precision, recall and F1 score, respectively. The promising results indicate that, even though our approach does not need any human-annotated labels, it outperforms the current state-of-the-art baseline trained with labeled data.

To sum up, the main contributions of this paper are threefold:
\begin{itemize}
    \item To our best knowledge, we are the first to design an unlabeled approach focusing on API recognizing and linking in unformatted text corpora. 
    \item We build an API contextual linker, making the model automatically link API mentions to an API repository, taking the sentence context into account.
    \item The experiment results show ARCLIN can discover traceability links between APIs and the repository more accurately, comparing with state-of-the-art baseline models. The code and dataset are released\footnote{Please find the resources in https://github.com/YintongHuo/ARCLIN.}.
\end{itemize}
 



\section{Problem Statement}\label{section:ProblemStatement}
In this section, we first introduce the main concepts used in this paper in Section~\ref{subsec:termonology} and then provide a formal definition of the task in Section~\ref{subsec:task-description}.

\subsection{Terminology}\label{subsec:termonology}
\begin{figure}[tb]
\centering
{\includegraphics[width=\linewidth]{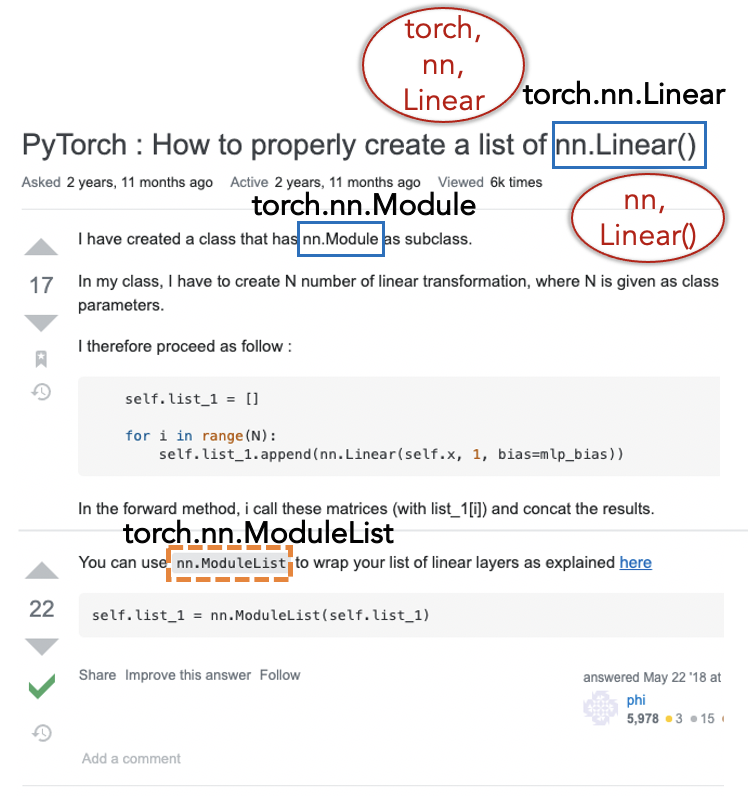}}
\caption{A screenshot of one StackOverflow post.}
\label{fig:terminology}
\end{figure}
API mining is the task of recognizing API mentions from free texts and linking the recognized API mentions to the corresponding API repositories. 
Figure~\ref{fig:terminology} is a screenshot of a StackOverflow post\footnote{The entire page is in https://stackoverflow.com/questions/50463975/pytorch-how-to-properly-create-a-list-of-nn-linear.}, we use this screenshot to illustrate the concepts used in this paper. 
Here, an \textit{API} could be a class name, a method name, or an attribute of a class. 
The term \textit{free text} (also called \textit{unformatted text}) refers to the text without any HTML tags (e.g., $<$code$>$). 
Orange dash box in Figure~\ref{fig:terminology} shows an example of $<$code$>$ usage.
An \textit{API mention} in texts is a token appearing in the free text that refers to a specific API in the repository. 
Blue box in the figure shows two API mentions (i.e., \textit{nn.Linear()} and \textit{nn.Module})\footnote{\textit{nn.ModuleList} in Orange box is also an API mention after removing the $<$code$>$ tag.}.
An \textit{API repository} is a collection of all entire qualified API names. \textit{Entire qualified name} is the exact API's name shown in its official website (e.g., \textit{torch.nn.Linear}, \textit{torch.nn.Module}). Each API's name in this repository is called an \textit{entry}. An entry is composed of sub-fields, splitted by ``.'', which are called entities, the entities of \textit{nn.Linear()} and \textit{torch.nn.Linear} are shown in nearby red circles.






\subsection{Task Description}\label{subsec:task-description}
Given a natural language sentence $S$ in free text and an API repository $D=\{D_1, D_2, ..., D_n\}$, where $D_i$ refers to an entry in the repository, the API mining task is to link API mentions to an entry in the API repository. 
The task involves two phases: (1) recognizing API mentions in the sentence; (2) linking API mentions to the corresponding entries in the API repository.
In practice, we first tokenize $S$ into a token list $[token_0, token_1,...,token_n]$, then $\forall{ 0\leq i \leq n}$, we determine whether $token_i$ refers to the element $D_j \in D$.

\section{Approach}\label{section:Approach}
\begin{figure*}[tb]
\centering
{\includegraphics[width=\linewidth]{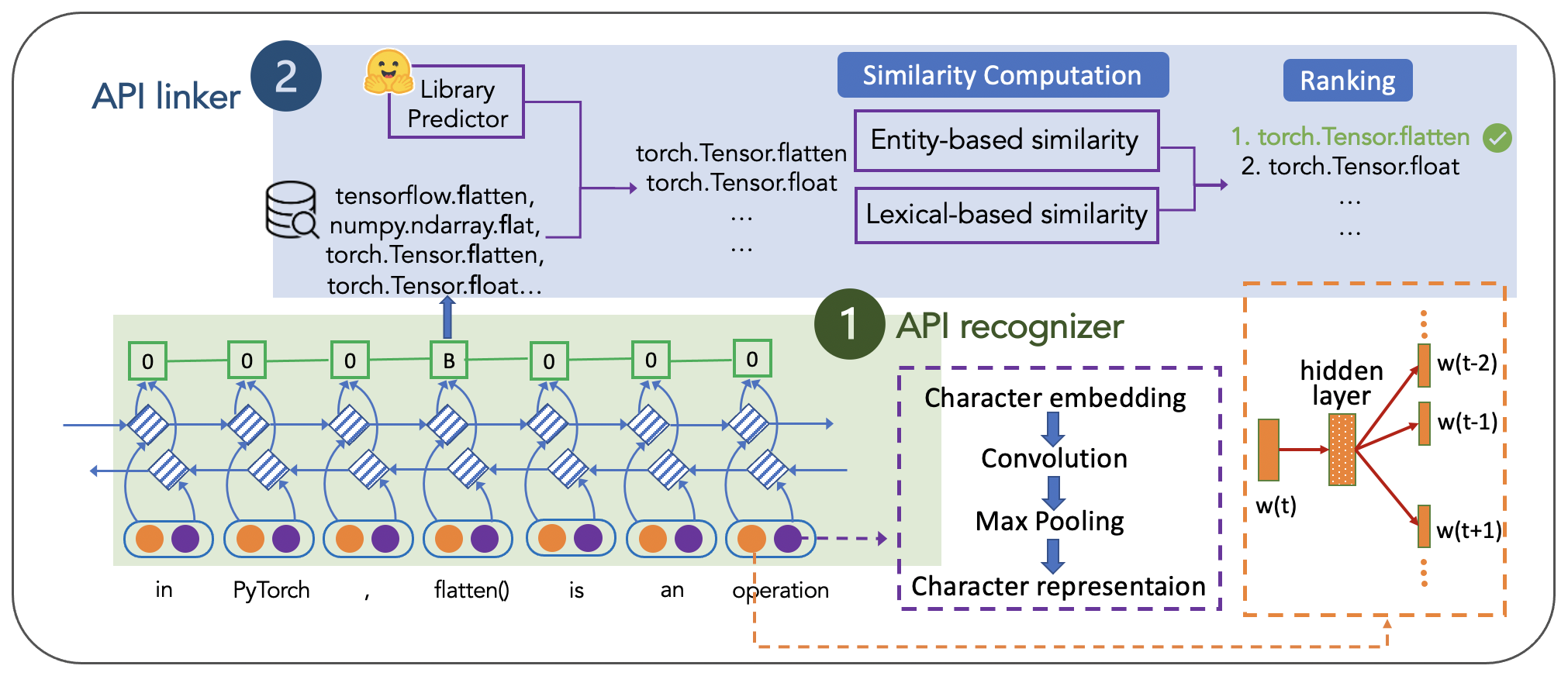}}
\caption{The framework of ARCLIN.}
\label{fig:framework}
\end{figure*}

In this section, we introduce our approach, including data preparation, an API recognizer, and a contextual API linker. Figure~\ref{fig:framework} shows the overall framework of ARCLIN. 
To begin with, sentences are fed into the API recognizer to uncover API mentions. Specifically, a context encoder is applied to acquire contextual embeddings of tokens by a bidirectional Long-Short Term Memory (LSTM) network, then these representations are decoded via a Conditional Random Field (CRF). The tokens which decoded as API mentions are sent to the API linker.
Next, an API linker is designed for discovering the most possible matched entry in the repository. To do so, we first generate a series of candidates by heuristic rules, then a library predictor narrows down the candidates by specifying a library. After that, we use an integrated scoring function to rank $<$mention, entry$>$ pairs. Finally, the candidate with the highest similarity above the threshold will be chosen as a link.

\subsection{Data Preparation}
\subsubsection{Text Corpus}\label{subsec:data-preprocess}
Given some libraries, we crawl all questions tagged with at least one of the given libraries from an online technical forum. 
Besides questions and answers, ~\citet{zhang2019reading} revealed that the majority of comments were also informative as they provided a supplementary view to the answer. Therefore, for each question-answering thread, we crawl the question, all answers, and their comments. We discard code snippets in $<$pre$>$ $<$code$>$ but keep contents in $<$code$>$ when it appears in a natural language sentence. 

StackOverflow users highlight API mentions in a natural language sentence by $<$code$>$ tags. However, ~\citet{tabassum-etal-2020-code} shows that 47\% of the code mentions are not indicated with this tag. 
If we only rely on the tags to do the mention detection, we will miss a large number of mentions.
Moreover, it is observed that contents in code tags can be noisy, many non-code elements (e.g., variables name, key points, or user name) are also highlighted by code tags~\cite{tabassum-etal-2020-code}.
To address this noisiness issue, as well as to generalize to other scenarios that cannot use $<$code$>$ tags (e.g., emails), we remove all $<$code$>$ tags in sentences and the markdown markers to make this task more similar to the real application. After collecting the data, we tokenize all sentences with the NLTK~\cite{bird2009natural} sentence parser. As a result, we obtain a set of parsed sentences in free text.

\subsubsection{Repository Construction}
For each given library, we crawl all APIs with their entire qualified names from official documentations. For instance, the APIs in the PyTorch library include methods (e.g., \textit{torch.Tensor.dim()}), functions (e.g., \textit{torch.nn.functional.avg\_pool1d()}), classes (e.g., \textit{torch.nn.AdaptiveAvgPool1d}), and attributes such as \textit{torch.backends.cudnn.enabled}. The API repository is made up of all crawled APIs' names.

\subsubsection{Tokenizer} 

We adapt a software-specific tokenizer used in~\citet{ye2016software} and~\citet{ye2016learning}, which preserves the integrity of an API mention. Current popular tokenizers such as SpaCy~\cite{spacy}, Stanford Parser~\cite{de2006generating}, and NLTK all parse \textit{numpy.shape()} into a token list of \textit{[``numpy.shape'', ``('', ``)'']}, but the deployed software-specific tokenizer will treat \textit{numpy.shape()} as a single token.

\subsubsection{Inverse Document Frequency (IDF)} IDF is a way to measure the importance of a word in a corpus. A word's IDF is disproportionate to the word's frequency. 
Given the assumption that if a word frequently occurs in a document, it may contain relatively less information, the formula for computing IDF for a word $w$ is shown in Equation~\ref{Equ:IDF}:
\begin{eqnarray}
IDF(w) = log(\frac{\#\text{Documents\_Number}}{\#\text{Document\_with\_w}+1})
\label{Equ:IDF}
\end{eqnarray}

In this paper, we compute two types of IDF, $IDF_{token}$ and $IDF_{entity}$. 
We use $IDF_{token}$ to measure the token's importance in a corpus, where we regard each sentence as a document. 
For $IDF_{entity}$, we compute the entity's importance in the repository. We consider each entry is a document and its entities are words. For example, the document \textit{numpy.reshape()} has two words: \textit{``numpy''} and \textit{``reshape()''}.
Intuitively, since all Numpy APIs contain the entity \textit{``numpy''}, its IDF value is relatively low.

\subsection{Recognizer}
The first step of mining APIs from the sentences is extracting API mentions without specifying which APIs they refer to.
At this step, we propose an automatic API mention recognizer that prefers recall over precision to cover as many API mentions as possible.
We first introduce an automatic approach to mine natural (but noisy) labels, then elaborate on architectures of the recognizer in the following subsections.
\subsubsection{Automatic labeling} 
Traditional machine learning approaches label a large-scale training set to train classification models for the task. However, considering that there are enormous APIs even in one programming language, it is infeasible to obtain sufficient human-labeled data for all of them. There is a need to devise an algorithm that escapes from any human annotation. 

Previous studies~\cite{sang2003introduction,ratinov2009design} show that prior external knowledge (i.e., API repository) was critical for good performance in identifying named entity in a sentence. Motivated by this, we use the following criteria (i.e., domain knowledge) to automatically annotate potential API mentions:
\begin{itemize}
    \item If a token is exactly the entire qualified name (i.e., same as an entry in API repository), we regard the token as an API mention.
    \item Inspired from that users usually use ``\textit{()}'' at the end of a token to represent an API method name or function name, we treat the token as an API mention if the token contains ``\textit{()}''.
    \item Users also use ``.'' (e.g., \textit{numpy.shape()}, or \textit{x.shape()}) when they mention an API; thus, we consider the token is an API mention if it contains ``.''. To distinguish such mentions from emoticon or punctuation, we require the token to consist of more than three characters.
\end{itemize}

Moreover, to address the common-word polysemy problem introduced in Section~\ref{section:introduction}, we employ a data augmentation technique for each sentence with at least one API mention being detected. Specifically, we randomly replace the originally detected API mention with a new one only containing the last part of the name (e.g., \textit{x.view} will be replaced with \textit{view}). This data augmentation process forces the recognizer to learn contextual information of an API mention.


The self-labeling process inevitably introduces some noisy labels. For instance, even if the token \textit{``python2.7''} consists ``.'', it is not an API mention; Besides that, a missing space between two sentences (e.g., ... plot 500 ellipses on a single graph.If you do ...) will generate the wrong label for the token \textit{``graph.If''}. However, the proposed contextual linker is able to mitigate these noisy labels.
After automatic labeling, we feed the self-labeled data as well as the augmented ones into our context encoder.

\subsubsection{Context Encoder}
Context encoder is responsible for acquiring contextual word embeddings in a sentence.
The long short-term memory (LSTM) network has shown promising results in sequential labeling tasks~\cite{sundermeyer2012lstm}, due to its strong ability to capture long-distance context information.
The memory unit in LSTM enables it to generate the representation based on both the short-distance and long-distance context.
In this work, we design an LSTM network to achieve the goal. A bi-directional LSTM (Bi-LSTM)\cite{graves2013speech} is specifically used for preserving both past and future information within a sentence.

The architecture of the constructed encoder is shown in Figure~\ref{fig:framework}, where two granular-level features are considered. By doing so, the Bi-LSTM encoder simultaneously grasps word-level semantics and character-level details. Firstly, word embedding techniques are used to extract word-level semantics. Word embedding represents words as distributed vectors in a low-dimensional space so that words with similar semantic or syntactic meaning tend to be close in their vector space. Assuming that words present in a similar context have similar meanings, the common approach Skip-gram (Word2Vec)~\cite{mikolov2013distributed} learns word embeddings by predicting surrounding words given the central word. Similar to previous research~\cite{fu2017easy, xu2016predicting,huang2018api}, we train domain-specific word embeddings by Skip-gram on a domain corpus.


Secondly, as previous study~\cite{ma2016end} has shown that character-level representation is crucial to extract morphological evidence, we use this feature to alleviate the second morphological problem mentioned in Section~\ref{section:introduction}. Besides, since developers write API mentions with customized variable names under different scenarios, deploying character-level embedding allows us to cope with unseen words, named the out-of-vocabulary (OOV) problem. In particular, we elicit character-level features from the architecture shown in blue-dotted rectangles in Figure~\ref{fig:framework}, which incorporates one max pooling layer after a Convolutional Neural Network (CNN) is applied.




\subsubsection{Tag Decoder}
Given contextual word representations in a sentence, the tag decoder is used to determine whether the word is an API mention or just a common word. Inspired by previous sequence labeling works~\cite{lafferty2001conditional} in the natural language processing domain, we adopt a Conditional Random Field (CRF) to conduct the tag decoder on top of the text encoder (i.e., Bi-LSTM layer). By accurately obtaining structural dependencies among adjacent words in a sentence, the CRF module jointly predicts the tag of each word sequentially instead of predicting tags independently.
In order to balance between API mention coverage and precision in predictions, we select $Top\_P$ paths with the highest confidence score as the result of the CRF layer. If the token in $K(0 \leq K \leq P)$ paths is predicted as an API mention, we treat the token as an API mention and feed it to the contextual linker.

\subsection{Contextual Linker}
Once we obtain the API mentions in the text, ARCLIN links the correct API mentions to an entry in the repository. The core idea behind this linker is a series of disambiguation methods. 
Specifically, we firstly select entries as candidates in the repository, then rank the similarity score of every $<$mention,entry$>$ pair with the help of the mention's context information. 
Although the predicted API mentions may contain errors, the wrong mention will be hard to find an entry with a high similarity score. From this aspect, the noise introduced by the last step will not affect the final results.
\subsubsection{Candidate Selection}
To reduce the time complexity of comparing all entries in the repository with the API mention, we narrow the scope by listing a set of candidates.
Inspired by the fact that, even though humans can make errors in spelling words, such misspelling is hardly seen at the beginning of the word. So do the developers. 
Given a mention, we directly compare its last part (i.e., last entity) and the last part of the entries in the repository. If the first two characters of the last entities are case-insensitive matching, we add the entry to a candidate list.

\subsubsection{Library Predictor}
As the third challenge aforementioned, similar API entries in different libraries bring difficulties to disambiguate the mention. An intuitive way is to take sentence-level semantics into consideration.
To capture rich contextual information from sentences, we first train the most popular language model BERT~\cite{devlin2019bert} with all training sentences for each library. Then, one fully connected layer followed by a soft-max output layer is fine-tuned to predict the library of input sentences based on the semantic embedding produced by BERT.

\subsubsection{Similarity Computation}
Given an API mention $m$ and its candidates $e$, we calculate the similarity score between the API mention and each candidate. Finally, we rank all candidates based on their similarity and select the most relevant candidate above the threshold.
Basically, we compute similarity based on bag similarity. Given two bags of entities, $M$, $E_i$ being split by ``.'' from the API mention $m$ and a candidate $e_i \in e$, respectively, we compute the similarity from two aspects, lexical-similarity and entity-similarity.

\textbf{Lexical Similarity.} 
This step is motivated by the fact that sometimes developers make spelling errors in sentences, especially when the mentioned API name is long. However, even if we make a typo in some words, its lexical meaning (i.e., word representation learned from corpus) will not change.

Inspired by~\citet{huang2018api}, we use the Equation~\ref{Equ:lexical-sim} to calculate lexical-based similarity between mention entities $M$ and entities of one candidate $E_i$.
\begin{eqnarray}
Sim_{L}(M\rightarrow E_i) = \frac{\sum_{w\in M}sim(w, E_i)*IDF_{token}(w)}{\sum_{w\in M}IDF_{token}(w)},
\label{Equ:lexical-sim}
\end{eqnarray}
where $IDF_{token}(w)$ represents the IDF value of token $w$ in the training data. $sim(w,E_i)$ refers to the maximum lexical similarity score between the element $w\in M$ and elements in set $E_i$. 
We calculate lexical similarity for pairs of entities by another word embedding model FastText~\cite{bojanowski2017enriching}.
Unlike Word2Vec, FastText is the embedding model that incorporates n-gram features of a token, so it solves the OOV problem.
Inversely, we also compute the similarity $Sim_L(E_i\rightarrow M)$ by exchanging $M$ and $E_i$ in Equation~\ref{Equ:lexical-sim-all}.
In the end, the overall lexical similarity is formulated through an arithmetic mean operation:
\begin{eqnarray}
Sim_L(M, E_i) = \frac{Sim_L(M\rightarrow E_i) + Sim_L(E_i\rightarrow M)}{2}.
\label{Equ:lexical-sim-all}
\end{eqnarray}

\textbf{Entity Similarity.}
Jaccard similarity coefficient~\cite{jaccard1912distribution} is widely used in gauging how similar the two sets are. Given two bags of entities $M$, $E_i$, we formulate our weighted Jaccard similarity as Equation~\ref{Equ:entity-sim}:
\begin{eqnarray}
Sim_J(M, E_i) = \frac{\sum_{w \in (M\cap E_i)}IDF_{entity}(w)}{\sum_{w \in E_i} IDF_{entity}(w)},
\label{Equ:entity-sim}
\end{eqnarray}
where $IDF_{entity}(w)$ represents the IDF value of entity $w$ in the API repository. IDF provides a standard to measure the salience of a token. a higher IDF value represents that it appears more frequently, carrying lower information entropy. 
For instance, in our repository, tokens such as \textit{nn}, \textit{torch}, \textit{numpy} contain a low IDF value since it is almost present in every entry, but tokens such as \textit{AdaptiveMaxPool1d} and \textit{binary\_cross\_entropy} deserve more attention, thus a high IDF value.
Intuitively, instead of class or module names, we always use method names to clarify the mentioned API, which contains a higher IDF value. Such discriminative tokens contribute significantly to this $Sim_J$ function, while missing a match in \textit{nn} just makes a minor effect on the entity similarity score.

\textbf{Overall Similarity.} To sum up, the scoring function for calculating similarity is composed of a lexical similarity function and an entity similarity function. Given an API mention $m$ and an entry $e_i$, the overall similarity is calculated by Equation~\ref{Equ:overall-sim}:
\begin{eqnarray}
Sim(m,e_i)=Sim_L(m,E_i) + Sim_J(m,E_i),
\label{Equ:overall-sim}
\end{eqnarray}
where $Sim_L(m,E_i)$ and $Sim_J(m,E_i)$ are defined above.
To exclude the API mentions that are wrong predictions introduced from the recognizer, and the API mentions that refer to an API out of our repository, we eliminate the candidates $e_i \in e$ with lower $Sim(m,e_i)$ value than the similarity threshold $S$. Finally, we rank all remaining candidates and choose the $e_j \in e$ with the highest $Sim(m,e_j)$ as output.

\section{Experimental Setup}\label{section:Setup}
In this section, we introduce the experimental setup details, including data collection, implementation details, and evaluation metrics.

\subsection{Data Collection}
\subsubsection{Text Preparation}
In this paper, we focus on five widely-used third-party libraries in Python: \textit{Pytorch}, \textit{Pandas}, \textit{Tensorflow}, \textit{Numpy}, \textit{Matplotlib}. We crawl all questions tagged with at least one of the above libraries using Scrapy in StackOverflow. For each question-answer thread, we collect questions, all answers and their comments. Details of the data preprocessing method are described in Section~\ref{subsec:data-preprocess}.

\subsubsection{API Repository}
We construct an API repository containing all API in five chosen third-party libraries with their entire qualified names. We use Scrapy to crawl all APIs from their official websites. Information such as the version of each library, the number of APIs in each library is listed in Table~\ref{tab:testing-data}. Considering that parentheses \textit{``()''} are not the sign to differ APIs from each other, we remove all \textit{``()''} at the end of the API entire qualified names (e.g., store \textit{numpy.einsum} instead of \textit{numpy.einsum()}).
\begin{table}[t]
\small
    \centering
        \caption{Statistics of API repository and Py-mention set.}
    \begin{tabular}{c||c|c|c|c}
    \toprule
    Library & Version & \#API & \#Mention & \#Sentence\\
    \midrule
        PyTorch~\cite{PyTorch} & 1.8.0 & 2,472 & 133 & 562\\
        Tensorflow~\cite{Tensorflow} & 2.4.1 & 10,361 & 87 & 532\\
        Pandas~\cite{Pandas} & 1.2.4 & 2,174 & 117 & 573\\
        Numpy~\cite{Numpy} & 1.20  & 1,913 & 116 & 580\\
        Matplotlib~\cite{Matplotlib} & 3.4.1 & 6,937 & 105 & 583\\
        
    \midrule
    Sum & - & 23,857 & 558 & 2,830 \\
         \bottomrule
    \end{tabular}
    \label{tab:testing-data}
\end{table}
\subsubsection{Dataset}
Considering all texts crawled from text preparation are too large to cope with, we randomly sample 150,000 sentences for each of the libraries and treat them as unlabeled training data. 
After applying self-labeling and data augmentation, we obtain 125,931 sentences for training the recognizer.
The distributions of training data, automatically API labels, and augmentation results are shown in Table~\ref{tab:training-stats}.
\begin{table}[t]
\small
    \centering
        \caption{Statistics of training set.}
    \begin{tabular}{c||c|c|c}
    \toprule
    Library & \#Sentence & \#Autolabel & \#Augmentation\\
    \midrule
        PyTorch & 150,000 & 11,057 & 27,077 \\
        Tensorflow & 150,000 & 8,925 & 21,899\\
        Pandas & 150,000 & 10,537 & 25450\\
        Numpy & 150,000 & 11,501 & 27,941\\
        Matplotlib & 150,000 & 9,769 & 23,564\\
    \midrule
        Sum & 750,000 & 51,789 & 125,931\\
         \bottomrule
    \end{tabular}
    \label{tab:training-stats}
\end{table}

For the testing data, we randomly select 600 sentences from each library (without overlapping with the training data) and ask experts to annotate them. To ensure annotation quality, two invited experts both have more than four years of experience in Python development and are all familiar with five libraries.
Considering that a long sentence is more likely to contain API mentions, we select the testing data sentences longer than ten tokens.
During the annotation, given the whole API repository, experts are asked to annotate whether each token in a sentence is referring to an API in the repository or not. If yes, they need to write down the entire qualified name of an API mention.  We also ask annotators to throw away the sentence if they are not confident at what it refers to.
In this way, we collect 2,830 sentences with 558 API mentions from five libraries in total, where their distributions along with the repository's distribution are shown in Table~\ref{tab:testing-data}. Typical examples are below:
\begin{itemize}
    \item If you don't want to export, please uncomment \textit{plt.show()} \Blue{\textit{[matplotlib.pyplot.show()]}} and remove ...
    \item I've usually gotten good performance out of numpy's \textit{einsum} \Blue{\textit{[numpy.einsum()]}} function and I like ... 
    \item Here is a way to do it using \textit{stack} \textit{\Blue{[torch.stack()]}} or \textit{unbind} \textit{\Blue{[torch.unbind()]}}.
\end{itemize}

Here, \textit{black italic fonts} indicates API mentions and \textit{\Blue{blue italic fonts}} in brackets are the linked APIs in the repository (with entire qualified names).


\subsection{Implementation Details}

In the data preprocessing period, we train a skip-gram Word2Vec model based on our corpus with gensim~\cite{rehurek_lrec}. We also train a FastText word embedding model with gensim~\cite{rehurek_lrec} for computing $<$mention, entry$>$ pair lexical-based similarity. The embedding size for Word2Vec and FastText models are set to 300. Two models are trained for ten epochs\footnote{Word2Vec and FastText models converge before ten epochs.}. $IDF_{token}$ and $IDF_{entity}$ are trained on the training data.

For the API recognition part, we use an open-source natural sequence labeling tool from~\cite{yang2018ncrf} as the implementation and train the recognizer on the augmented data. The character embedding size is set to 30, and the layer number of Bi-LSTM is set to one. We train the recognizer with the learning rate as 0.001 for five iterations. We choose five paths with the highest confidence score in the CRF layer, and treat a token as an API mention if and only if it is predicted so in at least two out of five paths ($Top\_P=5, K=2$).
For the API linker, we train our library predictor with Transformer~\cite{wolf-etal-2020-transformers} for ten iterations with the learning rate of 0.001. The default threshold $S$ for the scoring function is 1.1 unless we specify them with other values.





\subsection{Baselines}
To the best of our knowledge, there is no existing work focusing on extracting API links from unformatted texts. 
We compare our method with the following baselines: APIReal is the most relevant work to ours but they mine APIs from StackOverflow posts, and the other two baselines are rule-based.
\subsubsection{APIReal} ~\citet{ye2018apireal} proposed the model named APIReal, which predicted API recognition and linking in a StackOverflow post. APIReal contains two stages similar to ours: a recognizer to extract API mentions and a linker to link API to the repository. In the recognizer, they manually labeled the training data to learn API mentions by feeding human-crafted features into a CRF model. In the linker, 
they utilized external information, such as the question title, contents in the code block, $<$code$>$ tags, and \textit{URLs} in a post to predict what an API mention links to. 

When implementing this baseline, we ``counterfeit'' a file crawled from StackOverflow in the same input format, where each line is a sentence from our test set. 
In this way, APIReal will treat our file as a post from StackOverflow and continuously processes them. Moreover, as the database of APIReal includes three of five libraries comparing to ours (i.e., Pandas, Numpy, Matplotlib), we compute Precision, Recall, and F1 scores on the three libraries.
\subsubsection{RuleBase-Pure} We also include a pure Rule-based approach as the baseline. Specifically, we check whether each token in the sentence is the same as an entry in the API repository. This baseline provides us with insights into the quality of written API mentions in StackOverflow.
\subsubsection{RuleBase-Knowl} We also include a Rule-based approach with prior knowledge as a baseline. 
Here, prior knowledge refers to the common writing behaviors for API mentions in StackOverflow. 
Specifically, we replace \textit{``np''} with \textit{``numpy''}, \textit{``pd''} with \textit{``pandas''}, \textit{``tf''} with \textit{``tensorflow''} for each token, respectively. 

\subsection{Evaluation Metrics}
For fair comparison, we use \textit{Precision}, \textit{Recall}, and \textit{F1} scores to evaluate ARCLIN's performance in our test set, which is also used by all previous works~\cite{ye2018apireal,bacchelli2010linking, dagenais2012recovering}. Specifically, precision means what percentage of API linking predictions are correct, recall means what percentage of the real API mentions are covered, and F1 is the harmonic mean of precision and recall. 


\section{Experimental Results}\label{section:Result}
In this section, we discuss the performance of ARCLIN model by diving into three research questions from Section~\ref{subsec:rq1} to Section~\ref{subsec:rq3}: 

\textbf{(1) How effective is ARCLIN?}
\; We compare ARCLIN to three baselines in the proposed test set. The result shows that ARCLIN outperforms baselines by large margins, even though it is free from any labor-intensive annotations and handcrafted rules.

\textbf{(2) How effective are the components of ARCLIN?}
\; The devised framework is made up of an API recognizer and an API linker. The latter one includes a library predictor and a scoring function balance between the lexical similarity and entity similarity. 
To evaluate the contribution of each component, we discard each element at one time and implement the remaining part in our test set. Details of analysis are provided along with the experiment results.

\textbf{(3) What is the generalization ability of ARCLIN?}
\; Considering the large number of libraries in the real world, we are interested in how ARCLIN performs in mining APIs inside an unseen library. To explore its generalization ability, we train the model in one library and test it in another library.


\subsection{RQ1: How effective is ARCLIN?}\label{subsec:rq1}
ARCLIN aims to automatically extract API mentions from free text sentences and link them to an entry in the repository. Thus, to prove its effectiveness, we evaluate ARCLIN in sentences selected from StackOverflow posts. We feed test sentences into the ARCLIN model and examine whether it could mine correct APIs.

\begin{table}[t]
    \centering
        \caption{Experimental Results.}
    \begin{tabular}{l||c|c|c}
    \toprule
    Approach & Precision & Recall & F1\\
    \midrule
    RuleBase-Pure & 1.00 & 0.070 & 0.131 \\
        RuleBase-Knowl & 1.00 & 0.314 & 0.478\\
        APIReal & 0.787 & 0.604 & 0.683 \\
        \quad - w/o rules & 0.823 & 0.477 & 0.599 \\
        \textbf{ARCLIN} Ensemble & 0.784 & 0.742 & \textbf{0.762}\\
        \quad - PyTorch & 0.861 & 0.801 & 0.830 \\
        \quad - Tensorflow & 0.576 & 0.840 & 0.683\\
        \quad - Pandas & 0.717 & 0.778 & 0.746\\
        \quad - Numpy & 0.865  & 0.741 & 0.798 \\
        \quad - Matplotlib & 0.825 & 0.762 & 0.792\\
        
         \bottomrule
    \end{tabular}
    \label{tab:experimental-results}
\end{table}

To answer this question, we compare ARCLIN with a current state-of-the-art baseline named APIReal~\cite{ye2018apireal}, a purely rule-based approach RuleBased-Pure, and a rule-based approach incorporating prior knowledge named RuleBased-Knowl. 
Experimental results are shown in Table~\ref{tab:experimental-results}. Apart from the overall performance of ARCLIN in our whole test set (in ARCLIN Ensemble), we also examine the performance in every single library, shown in the following five rows. The results indicate our ARCLIN significantly outperforms all other baselines. From the results, we see that our ARCLIN model can achieve 78.41\%, 74.19\%, and 76.24\% in precision, recall, and F1 score, respectively.

It is worthy to notice that RuleBase-Pure only retrieves 6.99\% of all API mentions, reflecting that developers rarely write the entire qualified name when they mention some APIs. This is also one of the motivations of this work. 
The RuleBase-Knowl model provides better performance with the help of prior knowledge. However, if we want to extend the model to a large number of libraries, it is implausible for researchers to enumerate all possible abbreviations for each library. Although the model gives an acceptable performance, it can hardly be used extensively.
Another baseline APIReal reaches the F1 score of 0.683, which is lower than the performance in their dataset. We attribute the unfavorable performance to several reasons: 
(1) The constraints of handcrafted patterns in resolving customized variables. 
As illustrating in the second morphological challenge, the unprofessional developers usually write down API mentions with customized variables (\textit{a.reshape}) or aliases (\textit{np.reshape}). Since APIReal leverages a collection of pre-defined rules to solve the problem (e.g., \textit{np} for \textit{numpy}), the customized variables or uncommon aliases outside the scope lead to mistakes. The impact of such handcrafted rules is quantitated in the w/o rule line in Table~\ref{tab:experimental-results}.
(2) Difficulty in mining APIs in free texts. APIReal leverages $<$code$>$ tags in its recognizer; thus, once someone writes down API mentions in such tags, APIReals can easily extract them. But our dataset does not contain such signs to help the recognizer find out API mentions. 
(3) Insufficient information. APIReal utilizes information from source StackOverflow posts, such as \textit{URLs}, question titles, and code snippets. However, mining APIs from sentences in our task requires the model to capture a richer semantic meaning.

\begin{figure}[tb]
\centering
{\includegraphics[width=\linewidth]{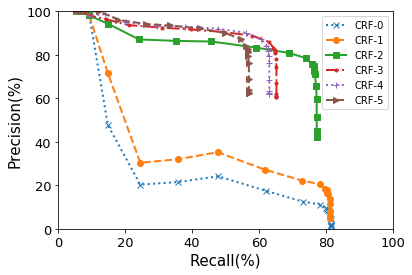}}
\caption{P-R curve of different hyper-parameters $K$ and $S$.}
\label{fig:PR-curve}
\end{figure}

\begin{table}[t]
    \centering
    \caption{Effectiveness of components in ARCLIN.}
    \begin{tabular}{l||c|c|c}
    \toprule
      & Precision & Recall & F1\\
    \midrule
    ARCLIN (ensemble) & 0.784 & 0.742 & 0.762 \\
    \quad - w/o recognizer & 0.112 & 0.783 & 0.195 \\
    \quad - w/o lib\_pred & 0.649 & 0.715 & 0.680\\
    \quad - w/o lexical\_sim & 0.814 & 0.439 & 0.570\\
    \quad - w/o entity\_sim & 0.645& 0.719& 0.680\\
    \bottomrule
    \end{tabular}
    \label{tab:Abalation}
\end{table}

\begin{table*}[t]
\small
    \centering
        \caption{The generalization ability of ARCLIN. P, R and F1 refers to precision, recall and F1 score respectively.}
\resizebox{\textwidth}{!}{
    \begin{tabular}{l||ccc|ccc|ccc|ccc|ccc}
    \toprule
    \multirow{3}{*}{Training}& \multicolumn{15}{c}{Testing}\\
    \cline{2-16}
      & \multicolumn{3}{c}{PyTorch} & \multicolumn{3}{c}{Tensorflow}  & \multicolumn{3}{c}{Pandas} & \multicolumn{3}{c}{Numpy} & \multicolumn{3}{c}{Matplotlib}\\
      & P & R & F1 & P & R & F1 & P & R & F1& P & R & F1 & P & R & F1\\
    \midrule
    PyTorch & - & -& -& 0.289 & 0.753 & 0.418 & 0.455 & 0.650 & 0.535 &0.472 & 0.741 & 0.576 & 0.852 & 0.657 & 0.742\\
    Tensorflow & 0.427 & 0.710 & 0.533 & -&-&- & 0.420 &0.701 & 0.526 & 0.422 & 0.768 & 0.544 & 0.738& 0.752 & 0.745\\
    Pandas & 0.472 & 0.634 & 0.541 & 0.284 & 0.753 & 0.412 & -&-&- & 0.469 & 0.741 & 0.574 & 0.833 & 0.667 & 0.741\\
    Numpy & 0.422 & 0.664 & 0.516 & 0.222 & 0.803 & 0.348 & 0.365 & 0.727 & 0.486 & -&-&- &0.817 & 0.724 & 0.768 \\
    Matplotlib & 0.449 & 0.641 & 0.528 & 0.268 & 0.765 &0.397 & 0.415 & 0.684 & 0.516 & 0.454 & 0.741 & 0.563 & -&-&-\\
    \bottomrule
    \end{tabular}
}
    \label{tab:Transferbility}
\end{table*}

ARCLIN reaches the highest performance among the three baselines. We conclude the reasons as follows: (1) ARCLIN owns the recognizer that keeps all possible API mentions by selecting the top five paths in CRF and conduct voting. In this way, ARCLIN will not miss too many API mentions; (2) ARCLIN's library indicator provides scope for library selection, preventing it from linking to the entry from wrong libraries; (3) ARCLIN's scoring function balances the lexical similarity and spelling similarity, so small variations of an API's name will not affect its final prediction.

In addition to the good performance, another advantage of ARCLIN is its flexibility. Figure~\ref{fig:PR-curve} provides a precision-recall curve to show how the performance is affected by the hyperparameters $K$ and $S$. Each curve \textit{CRF-K} in the figure represents a token will be considered as an API mention if $K (0\leq K \leq 5)$ out of five paths predict it so.
Each point in a curve is ARCLIN's performance under a similarity threshold $S(0 \leq S \leq 2)$.
A higher-scoring threshold means a matched $<$mention, entry$>$ requires a higher similarity.
Generally, a higher precision occurs simultaneously with a lower recall rate. 
ARCLIN is able to achieve 100\% precision under a low recall rate. Therefore, we can customize the threshold under different scenarios.
The figure also shows that we cannot achieve 100\% recall even the precision gets down to zero. We ascribe the situation into the following reason: Compared with character-disorder, word-disorder is too complex for ARCLIN to deal with. For example, when \textit{torch.nn.BCEWithLogitsLoss} is written as \textit{BCELosswithlogits}, even if ARCLIN narrows down candidates into the correct library, it is hard for ARCLIN to conduct API linking with each other.
To conclude, after evaluating ARCLIN in our test set the experiment result shows that it outperforms baselines by large margins, even though it is free from any labor-intensive annotations and handcrafted rules.

\subsection{RQ2: How effective are the components of ARCLIN?}\label{subsec:rq2}
ARCLIN is comprised of an \textit{API recognizer} and an API linker with a \textit{library predictor} and a scoring function balance between the \textit{lexical similarity} and \textit{entity similarity}.
To investigate the contribution of each module, we discard each component at a time, implement the model in our test set, and analyze its performance. Experimental results of this ablation study are shown in Table~\ref{tab:Abalation}, where each row below ARCLIN (ensemble) represents the result of a missing component. In the w/o lexical\_sim and w/o entity\_sim setting, we set the scoring threshold to 0.6.
Generally, the missing module negatively affects the model's performance more or less. We will discuss the effects in the following paragraphs respectively.

\subsubsection{NO Recognizer}
In this setting, the model tries to link an entry in the API repository for each token. The precision performance is dramatically decreasing because the majority of tokens in a sentence are not API mentions, but they can still be linked to an API in the repository because of their high similarity. 
For instance, the common word \textit{where} has a high similarity with the API \textit{numpy.where()} because both of them contain ``where'' within the token, but it is not an API mention. ARCLIN made lots of such mistakes, causing low precision.

\subsubsection{NO Library Predictor}
In this setting, the model tries to generate candidates from all five libraries, neglecting the sentence context information. The failure occurs when different libraries have a method with similar names. For instance, PyTorch has the method \textit{torch.stack()} while Numpy also contains the method \textit{numpy.stack()}, if a developer only writes \textit{``stack''} as the API mention, the model cannot disambiguate the token.

\subsubsection{NO Lexical Similarity}
Without the lexical similarity, the scoring function fully relies on the entity similarity. A $<$mention, entry$>$ pair will be linked if and only if some entities within them are exactly the same. This approach provides a high precision rate, since it is similar to an advanced rule-based algorithm. However, it cannot deal with spelling errors.
For example, \textit{np.zeros()} will be linked with \textit{numpy.zeros()} because both of them has the entity \textit{``zeros''}, but \textit{numpy.zeros()} cannot matched with \textit{np.zero()}, even if the API mention contains only one missing character.


\subsubsection{NO Entity Similarity}
In this case, the lexical similarity is determinative to the scoring function. This function works fine in most cases, but falling short when an API mention refers to a long API name. 
For instance, the API mention \textit{tf.layers.batch\_normalization} has a higher similarity score with \textit{tf.keras.layers.BatchNormalization()} rather than \textit{tf.compat.v1.layers.batch\_normalization()}. From a lexical perspective, \textit{batch\_normalization} is not far away from \textit{BatchNormalization}, so the final scoring function will easily be affected by other factors (i.e., missing module name in this example).

\subsection{RQ3: What is the generalization ability of ARCLIN?}\label{subsec:rq3}
Considering the large number of libraries even for one programming language, we are interested in the generalization ability of ARCLIN. A promising API mining model should have the ability to mine APIs without training on the library-specific corpus.

To answer the question, we train the recognizer and linker in one library corpus, then the model attempts to recognize and identify APIs of another library in our test set. In this setting, the model never sees the new library before, so the library predictor is removed from ARCLIN.

Table~\ref{tab:Transferbility} shows the generalization ability of each pair of libraries. The experimental results show ARCLIN gains the generalization ability to some extent. 
The experiments further indicates that the transferred model evaluated with Matplotlib achieves a higher performance. For instance, the model that has been trained from Numpy, is able to correctly recognize 81.7\% Matplotlib APIs, according to the Table~\ref{tab:Transferbility}.
We ascribe the reason to the distinctiveness of API's name in Matplotlib. Specifically, API names in Matplotlib (e.g., \textit{matplotlib.pyplot.pcolormesh()} are rather different from APIs in scientific computing libraries (e.g., \textit{numpy.zeros()} or \textit{torch.zeros()}), so the model is free from mistakenly linking to APIs in other libraries.


Generally, the transferred models contain a better recall rate rather than precision, and we discuss the reason as follows. Without library predictor, ARCLIN may link API mentions to the wrong library if they contain similar method names. For example, given a sentence ``I have trouble with concatenating a list of tensors using PyTorch's stack.'' where \textit{``stack''} here is labeled as \textit{torch.stack()} in ground truth during the testing phase. If we train the model in Pandas and evaluate its generalization ability in Numpy library, ARCLIN will link \textit{``stack''} to the API \textit{numpy.stack()}.
In summary, the experiment results demonstrate the effectiveness and robustness of generalization ability. Such library-transferred experiment mimics the real-world scenario of applying ARCLIN to mine APIs from unseen libraries.

\section{Case Study}\label{section:case-study}
In this section, we dive into three cases to specify why ARCLIN outperforms APIReal and Rule-K (i.e., RuleBase-Knowl), where ground-truth is shown in blue italic font. The experimental results of four API mentions (in blue) are presented in Table~\ref{tab:case-study} with respect to the two phases (i.e., API Recognizer and API Linker\footnote{API Recognizer is denoted as Recog and API Linker is denoted as Link for space limitation.}). One API mention is successfully identified and resolved if and only if the ``Link'' phase gives the correct answer (\checkmark). 

\begin{itemize}
    \item Is there a more convenient alternative to \textit{figure.add\_suplot} \Blue{\textit{[matplotlib.figure.Figure.add\_subplot()]}} if I have multiple figures ...
    \item You may try ticking the major axis using \textit{ax.set\_major\_locator} \Blue{\textit{[matplotlib.axis.Axis.set\_major\_locator()]}}  called with  \textit{ticker. MultipleLocator()} \Blue{\textit{[matplotlib.ticker.MultipleLocator]}}.
    \item If you don’t want to export, please uncomment \textit{plt.show()} \Blue{\textit{[matplotlib.pyplot.show()]}} and remove …
\end{itemize}
\begin{table}[t]
\small
    \centering
    \caption{Case study.}
    \begin{tabular}{l||c|cc|cc}
    \toprule
    Approach  & \multicolumn{1}{c}{Rule-K} & \multicolumn{2}{c}{APIReal} & \multicolumn{2}{c}{ARCLIN}\\
    API Mention  & - & Recog & Link & Recog & Link \\
    \midrule
    figure.add\_suplot & \xmark & \xmark & \xmark & \cmark & \cmark\\
    ax.set\_major\_locator & \xmark & \cmark & \xmark & \cmark & \cmark\\
    ticker.MultipleLocator & \xmark & \cmark & \cmark & \cmark & \cmark\\
    plt.show() & \cmark & \cmark & \cmark & \cmark & \cmark\\
    \bottomrule
    \end{tabular}
    \label{tab:case-study}
\end{table}


Compared ARCLIN with Rule-K and APIReal, we categorize the characteristics for three approaches. 
Firstly, Rule-K can only resolve the API mention with the qualified name or a collection of specific abbreviations, depending on the handcrafted rules. For instance, if we add the common writing behavior that a developer usually calls \textit{matplotlib.pyplot} by its alias \textit{plt}, Rule-K will try to replace the alias with its original name for each token, then find if the new token matches a fully qualified API name in the repository.
Secondly, we observe that APIReal is more flexible than rule-based matching algorithm, by uncovering some API mentions by the recognizer (e.g., \textit{ax.set\_major\_locator} and \textit{ticker.MultipleLocator}), allowing it to address the first common-word ambiguity challenge. Nevertheless, its API Linker is not perfect to resolve the ambiguity introduced by morphological mentions, mainly comes from the customized name, such as \textit{ax} or \textit{ticker}. APIReal detects aliases by handcrafted patterns (e.g., \textit{pd} for \textit{pandas}), thus the alias that is not covered by rules will be inappropriately coped with.
Last but not least, the cases demonstrate the effectiveness of ARCLIN. The carefully devised API recognizer enables it to detect API mentions in unformatted text. Besides, the API Linker with entity similarity forces the model to pay attention to the informative entities (e.g., \textit{set\_major\_locator}), and the lexical similarity allows it to address the misspelling in API mentions. Therefore, ARCLIN can even resolve the \textit{figure.add\_suplot} to \textit{matplotlib.figure.Figure.add\_subplot()} even if the mention leaves out the letter ``b''.



\section{Threat to Validity}\label{section:Threat}
In this section, we discuss three potential threats to the validity of ARCLIN and provide our solutions to alleviate these threats. 
The first one is the potential bias brought by manual annotation of the data. We evaluate ARCLIN the Py-mention dataset, which is annotated by two different annotators. To overcome the human bias and ensure the data quality, we not only employ domain experts instead of crowd-sourcing workers, but also throw away the sentences with uncertainty. Annotation examples and guidelines are provided at first. As a result, the annotators fully understand what they need to do and keep confidence in their annotation.

The second one is the limited recall rate. As shown in Figure~\ref{fig:PR-curve}, the recall cannot achieve 100\% regardless of the threshold. In other words, ARCLIN cannot cover all ground-truth labels. We owe this recall limitation to the reasons of observed word disorder in mentions. ARCLIN computes similarity based on lexical-level and entity-level, but it fails in comparing $<$mention, entry$>$ pairs in word disorder. 
For example, if we use \textit{BCELosswithlogits} to represent \textit{torch.nn.BCEWithLogitsLoss}, the similarity score from ARCLIN is close to \textit{torch.nn.BCELoss}, therefore, the final output gets perturbation by other factors. To alleviate the issue, an n-gram based similarity can be used to extend our ARCLIN model.

The third threat is style constraint. Currently, we evaluate ARCLIN with five Python libraries and achieve promising performance, but if we migrate the model to other programming languages, the inconsistency of function calling format will introduce this threat. For instance, in C++ language, we use double colon ``::'' to call a static function or declare the namespace identification. Besides, to call a function in a class, one may use ``-$>$'' from a pointer or use the node ``.'' from a C++ entity. ARCLIN uses ``.'' to split the API's entire qualified name into a bag of package entities for similarity computation. If we implement ARCLIN in another language (e.g., C++), it is necessary to implement new split marks.

\section{Related Works}\label{section:RelatedWork}

\textbf{API Recognition.} If we want to link an API to some other source, the first step is to recognize APIs. In this paper, we use a recognizer to recognize APIs in free text sentences.~\citet{dagenais2012recovering} adopted partial program analysis (PPA) to parse Java snippets and then extracts code-like terms in informal discussions.The difference between theirs and ours is, our paper targets extracting APIs from natural language sentences, but the above studies were about extracting APIs from code blocks (written in free texts).
~\citet{bacchelli2010linking} employed a rule-based approach to extract API mentions from e-mails by designing different regular expressions applicable to different languages.~\citet{treude2016augmenting} suggested different regular expressions for question and body to extract API mentions from StackOverflow posts.~\citet{rigby2013discovering} used island grammars to identify code elements from free text with the help of compound camel cased terms while ignoring the common-word ambiguity.
The most relevant research to us is APIReal~\cite{ye2018apireal}, but their approach was applicable to recognizing APIs from StackOverflow posts with $<$code$>$ tags, which was much easier than our setting.
Besides, instead of linking such fine granularity APIs, researchers also explored linking between textual documents and code artifacts for maintenance. Some works~\citet{antoniol2002recovering,marcus2003recovering,marcus2005recovery,chen2010extraction} used information retrieval (IR) techniques or leverage Latent Semantic Indexing (LSI) to recover traceability links between elements in natural language documentation and source code in software systems. These studies were different from this paper since they performed a coarse granularity linking.

\textbf{API linking.}
``Linking'' can refer to linking code artifacts to documents, or linking APIs from free-texts to its entire qualified name in the repository. Regarding to linking code artifacts to documents, ~\citet{bacchelli2010linking} used two string-match information retrieval techniques (i.e., vector space model and LSI) to link detected APIs from e-mails to source code artifacts.
The latter category is what we have done in this paper, the main idea of matching APIs with their entire qualified name is how to conduct the disambiguation.
~\citet{dagenais2012recovering} suggested a set of filtering heuristics to disambiguate the API mentions.~\citet{ye2018apireal} disambiguated API mentions in a StackOverflow post by utilizing information in code blocks, question titles, and the location where \textit{URLs} points to. The first paper did not address the common word polysemy, while the second research mitigated the morphological challenge by labor-intensive rules, which was different from ours.

\textbf{Mining Technical Forums.}
Nowadays, many researchers devote themselves to mining knowledge from technical forums (e.g., StackOverflow) to facilitate developers in their programming issues. For example, a popular scenario is API recommendation~\cite{xie2020api,huang2018api,rahman2016rack}, these papers suggested a list of API classes for a natural language query by mining StackOverflow posts. Specifically, given a natural language query,~\citet{huang2018api} firstly searched the most relevant 50 questions and extracting APIs from posts. Then, it ranked all candidate APIs by considering the query-title similarity and title-APIs similarity.
~\citet{li2018api} proposed another application that explores API caveat in such a technical forum and presented a system to help developers to tackle the problem of negative usage of APIs.
It is noticed that many works studied in StackOverflow talks about APIs, our work serves as a foundation of this work for facilitating them to recognize and identify the APIs without the entire qualified name.

\section{Conclusion}\label{section:Conclusion}
In this paper, we propose a novel framework ARCLIN for recognizing API mentions from free text and linking to an API repository.
ARCLIN is composed of two components, an API recognizer and an API linker. The API recognizer extracts API mentions from free texts and the API linker disambiguates the API mentions by a library predictor to address reference ambiguity, and a scoring function incorporating lexical similarity and entity similarity.
After training the model in an unlabeled StackOverflow corpus, we implement ARCLIN in a human-annotated dataset named Py-mention, the experimental results demonstrate that it significantly outperforms all baselines. Moreover, the experiment about generalization ability demonstrates that ARCLIN can extract APIs from a new library even though ARCLIN is trained from another libraries. 

\section{Acknowledgement}
The work was supported by the Guangdong Key Research Program (No. 2020B010165002) and the Research Grants Council of the Hong Kong Special Administrative Region, China (CUHK 14210920).

\balance
\bibliographystyle{ACM-Reference-Format}
\bibliography{sample-base}

\end{document}